# Exciton physics and device application of two-dimensional transition metal dichalcogenide semiconductors

Thomas Mueller[1] and Ermin Malic[2]

[1] *Institute of Photonics, Vienna University of Technology, Gußhausstraße 27-29, 1040 Vienna, Austria*

[2] *Department of Physics, Chalmers University of Technology, SE-412 96 Gothenburg, Sweden*

Corresponding authors: thomas.mueller@tuwien.ac.at, ermin.malic@chalmers.se

**Two-dimensional group-VI transition metal dichalcogenide semiconductors, such as $MoS_2$, $WSe_2$ and others, exhibit strong light-matter coupling and possess direct band gaps in the infrared and visible spectral regimes, making them potentially interesting candidates for various applications in optics and optoelectronics. Here, we review their optical and optoelectronic properties with emphasis on exciton physics and devices. As excitons are tightly bound in these materials and dominate the optical response even at room-temperature, their properties are examined in depth in the first part of this article. We discuss the remarkably versatile excitonic landscape, including bright, dark, localized and interlayer excitons. In the second part, we provide an overview on the progress in optoelectronic device applications, such as electrically driven light emitters, photovoltaic solar cells, photodetectors and opto-valleytronic devices, again bearing in mind the prominent role of excitonic effects. We conclude with a brief discussion on challenges that remain to be addressed to exploit the full potential of transition metal dichalcogenide semiconductors in possible exciton-based applications.**

**INTRODUCTION**

Besides graphene a plethora of two-dimensional (2D) materials exist [1], with a wide range of different physical properties and potential applications. Among them, one important subclass is the family of transition metal dichalcogenides (TMDs) with chemical formula $MX_2$, where M is a transition metal and X a chalcogen atom. Some materials within the TMD family have been identified as semiconductors with band gaps in the infrared and visible spectral regions and strong light-matter coupling, making them potentially suitable for a variety of applications in optics and optoelectronics. In their bulk form, TMDs have been studied for decades [2, 3], but the properties of 2D layers with atomic scale thickness differ dramatically from the bulk material characteristics, being one of the reasons for renewed interest in this material class.

In this article, we review the optical properties of TMDs and their applications in optoelectronic devices. We shall restrict ourselves to group-VI TMDs (M = Mo, W; X = S, Se, Te) in the semiconducting 2H phase, as these are the most studied and understood. The 2D character and weak dielectric screening give rise to enhanced Coulomb interactions, resulting in the formation of strongly bound electron-hole pairs (excitons) upon optical excitation. Excitons dominate the optical and optoelectronic response, even at room-temperature, and their properties are thus surveyed in the first part of this review article. In the second part, we provide an overview on the recent progress in optoelectronic device applications, again with a focus on the prominent role of excitonic effects. We conclude with a brief discussion on remaining challenges.

Despite being a rather young research field, much progress has been made in the past few years in the understanding of the exciton physics in atomically thin layers, particularly monolayers. Fig. 1 shows some important milestones on a timeline. In 2005, just one year after the first mechanical exfoliation of graphene [4], K. Novoselov, A. Geim and coworkers showed [5] that the same technique can be used to isolate other monolayers, including a 2D semiconductor – molybdenum disulphide ($MoS_2$). Most of the 2D materials research in the following years focused on the unique properties of graphene, until 2010 when T. Heinz and F. Wang demonstrated independently that a $MoS_2$ monolayer exhibits a direct band gap and strong photoluminescence (PL) emission [6, 7]. Already early on, it was recognized that the optical response of TMDs arises from excitons, but it took a few more years until quantitative measurements of the binding energy were reported [8-10]. In parallel, other exciton complexes, such as charged excitons (trions) [11-13] and biexcitons [14, 15], have been studied. The field of TMD-based valleytronics can be traced back to 2012, when valley-selective optical excitation was first demonstrated [16-20]. The advent of Van der Waals (VdW) heterostructures [21] since 2013, permitted the fabrication of TMD heterostructures and the observation of interlayer excitons [22, 23]. More recently, the role of dark exciton states has become a major field of research [25-27], as well as the emission of quantum light from localized states [28-32]. The research on the fundamental optical properties of TMDs has been paralleled by the development of optoelectronic devices. The structure of many of these resembles that of the first $MoS_2$ monolayer field-effect transistor, demonstrated in 2011 by A. Kis and colleagues [33]. A year later the same device structure was used to realize atomically thin phototransistors [34, 35]. Fundamental photoconductivity studies on TMD bulk [2] and monolayers [6] date back earlier. Atomically thin p-n junctions were first realized in 2014, both in lateral [36-38] and vertical [39-41] geometries, which allowed for the realization of light-emitting diodes, photodiodes, and solar cells. The demonstration of near-unity photon quantum yield in a TMD monolayer [42] in 2015 has opened the door for the development of highly efficient optoelectronic devices.

**EXCITONS IN TMD SEMICONDUCTORS**

TMDs and related VdW heterostructures present a new paradigm for fundamental exciton physics. Weak dielectric screening and strong geometrical confinement give rise to an extraordinarily strong Coulomb interaction resulting in fascinating many-particle phenomena, such as the formation of different types of excitons including optically allowed and forbidden dark excitons (Fig. 2a), as well as spatially separated interlayer exciton states (Fig. 2b) [43-48]. They exhibit binding energies in the range of 0.5 eV, which is one to two orders of magnitude larger than in conventional materials, such as GaAs [8-10, 49]. Thus, excitonic features are stable at room temperature and dominate the optical response and non-equilibrium dynamics of these materials. Also higher-order excitonic quasiparticles, such as trions [11-13] and biexcitons [14, 15], have been observed in monolayer TMDs.

The research field of TMD optics and optoelectronics experienced a boom in 2010 after T. Heinz [6] and F. Wang [7] showed that molybdenum disulfide ($MoS_2$) undergoes a transition from an indirect-gap semiconductor in the bulk case to a direct-gap material in monolayer

MoS$_2$. This results in a drastically enhanced PL (Fig. 3a). Analyzing the orbital character of the electronic states at the relevant high-symmetry points by first-principle calculations, one can show that the observed transition can be understood as a consequence of a momentum/orbital selective interlayer splitting of the main relevant energy levels [50].

The optical spectra of TMDs are characterized by the appearance of two strongly pronounced resonances that are denoted as A and B excitons [6]. Their origin can be traced back to the strong spin-orbit coupling in these relatively heavy materials lifting the spin degeneracy of the valence and the conduction band. While the splitting is relatively small for the conduction band, the valance band separation reaches values of approx. 200 meV in molybdenum-based and even 400 meV in tungsten-based TMDs [51]. As a result, two optical transitions are possible involving holes in the upper and lower energy spin valence bands.

**Excitonic binding energy.** There are different experimental techniques allowing the measurement of the excitonic binding energy. The latter corresponds to the difference between the renormalized free-particle bandgap energy $E_0$ and the absolute energy of the exciton transition $E_X$. Using scanning tunneling spectroscopy, $E_0$ was measured and combined with PL measurements of $E_X$ it resulted in an exciton binding energy of 0.55 eV for MoSe$_2$ monolayer supported on bilayer graphene [9]. Furthermore, similarly to the Rydberg series in the hydrogen atom higher excitonic transitions with decreasing oscillator strength appear in optical spectra of TMDs below the free-particle bandgap [8]. Reflectance contrast measurements on monolayer WS$_2$ revealed the position of bright s-like exciton states up to 5s allowing an extrapolation of $E_0$. The excitonic binding energy of the ground state could then be estimated as 0.3 eV for monolayer WS$_2$ on a SiO$_2$ substrate. The excitonic states in TMDs considerably deviate from the hydrogen Rydberg series regarding the energy spacing of the first states due to the non-local dielectric screening associated with the inhomogeneous dielectric environment of TMDs. Finally, the exciton binding energy could also be determined by exploiting the selection rules in one- and two-photon spectroscopy. While the first addresses s-type excitonic states, the latter accesses p-type states within the Rydberg-like series [9, 10]. Combining the data, one can again extrapolate the position of $E_0$ and estimate the exciton binding energy of approx. 0.55 eV for monolayer MoSe$_2$ on graphene [9] and 0.37 eV on WSe$_2$ on a Si substrate covered by a SiO$_2$ layer [10].

Similar information on the position of higher excitonic states could also be obtained through measurements of second-harmonic generation [49] or internal excitonic 1s-2p transitions [46]. Recent studies show that one can tune the exciton binding energy by hundreds of meV by engineering the surrounding dielectric environment [52, 53]. One can even design in-plane dielectric heterostructure with a spatially dependent binding energy [52]. Furthermore, strain can be used as a tool to control the optical properties of 2D materials. Recent studies have shown that excitonic resonances significantly shift towards lower energies in presence of tensile strain [54, 55]. Note that this is due to a decrease of the electronic band gap, whereas the excitonic binding energy shows only slight changes [56, 57].

**Excitonic oscillator strength and linewidths.** TMDs absorb more than 15% of incoming light at the lowest excitonic resonance energy [58] despite their atomically thin character. Considering their thickness of less than 1 nm, they achieve one order of magnitude larger sunlight absorption than GaAs and Si [59]. The linewidth of excitonic resonances is given by the dephasing time of the excitonic polarization, which is determined by radiative and non-radiative scattering channels [60]. Optical 2D Fourier transform spectroscopy allows to unambiguously separate the intrinsic homogeneous broadening from impurity-dominated inhomogeneous contributions. The measurement at low temperatures revealed a radiative lifetime of excitonic transitions of approximately 200 fs, corresponding to a linewidth of 3 meV for monolayer $WSe_2$ on a sapphire substrate [61]. In hexagonal boron nitride (hBN) encapsulated samples PL linewidths down to 1.7 meV have been observed [62, 63]. Furthermore, a linear increase of the excitonic linewidth for temperatures up to 50 K was observed due to non-radiative scattering. In a weak excitation regime, exciton-phonon scattering is the crucial mechanism. The linear increase with temperature $T$ is a signature of acoustic phonons, since the Bose distribution scales linearly with $T$ at small temperatures [60]. Optical phonons become important above 100 K and result in a super-linear temperature dependence resulting in a broadening of approximately 30-40 meV at room temperature. While for molybdenum-based TMDs intravalley scattering channels around the K valley are predicted to be crucial, for tungsten-based TMDs scattering into energetically lower dark excitons turns out to become dominant [60].

**Exciton valley polarization.** Large spin-orbit coupling in combination with circular dichroism allows a spin- and valley-selective excitation of excitonic states in TMDs. Here, right-hand or left-hand circularly polarized light induces optical transitions only in the K or the K' valley, respectively. This was theoretically predicted [16] and experimentally demonstrated in PL measurements [17-20, 64-68]. As a result, besides spin and charge, valley can be considered as a new degree of freedom holding promising application potential in novel valleytronic devices. To reveal the elementary mechanisms behind the generation and decay of the valley polarization, a number of time-resolved experiments were performed [43]. The valley polarization was observed to decay on the timescale of a few picoseconds. This was surprising, as the decay requires intervalley scattering with a large momentum transfer combined with an electron and hole spin flip process, which is expected to occur on a longer time scale [69]. However, there is an efficient exciton intervalley coupling mechanism via Coulomb exchange interaction [45, 70]. It couples resonant excitonic states in K and K' valleys giving rise to a decay of valley polarization on a picosecond time scale. A recent work suggests that the spectral position of dark exciton states plays an important role for the degree of valley polarization via intra-valley dark-bright scattering [71]. The Coulomb exchange coupling only occurs as a second-order process, since it requires a non-zero center-of-mass momentum that first has to be generated by exciton-disorder coupling, biexcitonic excitations, or exciton-phonon coupling. Recently, a direct Coulomb-driven intervalley coupling mechanism has been proposed [72]. This interaction resembles the Dexter coupling between two spatially separated systems, however, now coupling different valleys in momentum space. It gives rise to an efficient intervalley transfer of coherent exciton populations between A and B excitonic states with the same spin in different valleys. It suggests that the valley polarization vanishes and is even inverted for A excitons when the B exciton is resonantly excited. This theoretical prediction is supported by energy- and valley-resolved femtosecond pump-probe experiments and also provides an explanation for the recently measured up-conversion in PL [73].

**Exciton dynamics and photoluminescence yield.** A number of theoretical and experimental studies have been performed addressing the exciton dynamics in TMDs [43, 44]. The main steps during the relaxation dynamics are optical excitation, exciton formation, exciton thermalization and exciton decay. Optical excitation generates an excitonic polarization, which is often referred to as coherent excitons [74]. In the low excitation limit, they decay via radiative coupling and exciton-phonon scattering. An efficient phonon-assisted polarization-to-population transfer leads to the formation of incoherent excitons on a sub-picosecond time scale [75]. The following exciton thermalization driven by emission and absorption of acoustic and optical phonons results in a thermalized exciton distribution. This occurs through a cascade from optically excited higher excitonic states via the Rydberg-like series of bright and dark excitonic states towards the ground state [76]. In medium to strong excitation regimes, exciton-exciton scattering also becomes important, where in particular Auger scattering has a strong impact resulting in a rapid exciton-exciton annihilation [76]. Only a fraction of incoherent excitons located within the light cone with a vanishing center-of-mass momentum can decay via radiative recombination and contribute to the PL. The PL yield strongly depends on the temperature and on whether the energetically lowest exciton is an optically forbidden dark state [73, 77]. Generally, the PL yield in TMDs is low (< 1%), presumably due to defect states where excitons become trapped before they can recombine radiatively. It was recently shown, though, that in the case of $MoS_2$ a chemical treatment with an organic superacid leads to a high PL yield of 95% [42]. The underlying mechanism is assumed to be based on shielding the defect states. Furthermore, one can use strain as a tool to change the relative spectral position of excitons and this way tune the coupling between dark and bright excitonic states [78].

**Dark excitons.** Recent experimental and theoretical studies have shown that besides the regular bright excitons, optically inaccessible dark excitonic states (Fig. 2a) play a significant role for understanding the optical fingerprint and the dynamics in TMDs [25-27, 48, 74, 77, 79-81]. We distinguish between spin- and momentum-forbidden dark excitons (Fig. 2a). Their existence was indirectly demonstrated in temperature-dependent PL experiments showing an increased PL yield at higher temperatures in tungsten-based TMDs [25] – a behavior that can be explained by the presence of dark excitons that are energetically lower than the bright state. Here, most excitons thermalize in these dark states, which drastically reduces the efficiency of light emission [74]. The higher the temperature, the more excitons occupy the energetically higher bright states within the light cone resulting in the experimentally observed increased PL.

**Spin-forbidden dark excitons.** The spin-orbit interaction lifting the spin degeneracy of the conduction band has a different sign for tungsten- and molybdenum-based TMDs [51]. As a consequence, while in $MoS_2$ and $MoSe_2$ spin-allowed bright excitons have the lowest energy, in $WS_2$ and $WSe_2$ the spin-forbidden states consisting of Coulomb-bound electrons and holes with opposite spin (Fig. 2a) are the energetically lowest – in excellent agreement with the measured temperature-dependent PL yield [77]. Brightening of these states was first demonstrated in low-temperature magneto-PL experiments, where strong in-plane magnetic fields were applied [25, 26, 44]. The latter couple to the charge carrier spin via the Zeeman effect, which mixes the spin-split bands and softens the spin-selection rule. As a result, dark excitons appear as clear features approximately 50 meV below the bright exciton resonance in magneto-PL spectra of tungsten-based TMDs (Fig. 3b). The peak shows a doublet structure that is ascribed to the splitting of the spin-forbidden states due to the Coulomb exchange coupling. Recently, spin-forbidden dark excitonic states could also be activated in an alternative approach through coupling to surface plasmon modes [27], which involves only placing $WSe_2$ monolayers on patterned silver surfaces. Dark excitons in $WSe_2$ have a non-zero dipole moment normal to the material plane allowing them to couple to surface plasmon modes.

**Momentum-forbidden dark excitons.** Intervalley dark excitons with Coulomb-bound electrons and holes located at different valleys in the momentum space present important scattering channels for exciton-phonon processes. In particular, the dark KΛ excitons with the hole located at the K and the electron at the Λ valley (Fig. 2a) are of great interest in tungsten-based TMDs, where they are believed to be the energetically lowest ground state. A recent scanning tunneling spectroscopy study suggests that the Λ valley is about 80 meV below the K valley in monolayer $WSe_2$ [82] suggesting that the momentum forbidden KΛ excitons will be energetically lower than the bright KK states. Furthermore, strain-dependent PL measurements evaluating the relative spectral changes between the K and the Λ valley indicate an opposite behavior for $MoS_2$ and $WSe_2$ implying that the latter is an indirect gap material [83]. A joint theory-experiment study investigating the homogenous linewidth of excitonic resonances comes to the conclusion that in tungsten-based TMDs scattering of optical and acoustic phonons into the energetically lower lying KΛ excitons leads to a considerable broadening even at very small temperatures [60]. Furthermore, the linewidth of excitonic resonances in TMDs can be controlled with strain. While the A exciton substantially narrows and becomes more symmetric for the selenium-based monolayer materials, no change is observed for WS$_2$ and even a linewidth increase is found for MoS$_2$ [84]. These effects can be traced back to the strain-induced changes in the exciton-phonon scattering due to the modification of the spectral distance between bright and dark excitons. Furthermore, indications for phonon-assisted radiative recombination pathways to momentum-forbidden states have been found in recent PL measurements [79]. A recent theoretical work predicts the activation of these excitons in presence of high-dipole molecules [80]. The latter disturb the translational symmetry of the TMD lattice and can provide the required momentum to reach the KΛ excitons – depending on the molecular dipole moment and molecular coverage. The efficient exciton-molecule coupling results in the appearance of a new exciton transition (Fig. 3c) that could be exploited for a dark-exciton-based detection of molecules. Finally, the presence of momentum-forbidden dark states could be experimentally revealed by probing the intra-excitonic 1s-2p transition [81]. Distinguishing the optical response shortly after optical excitation and after exciton thermalization allowed to demonstrate both in experiment and theory the transition of bright exciton population into the energetically lowest momentum-forbidden dark excitonic states in monolayer WSe$_2$.

**Localized excitons.** Due to the presence of impurities and/or strain in realistic TMD samples, electrons and holes can be trapped in potentials resulting in localized excitons. While at room temperature the PL spectra of TMDs are determined by a broad peak corresponding to the bright 1s exciton transitions, at low temperatures a series of additional spectrally narrow resonances appears below the bright 1s excitons (Fig. 4a) [85]. These resonances can be ascribed to free excitons trapped in local potential wells. They vanish at higher temperatures, where the thermal energy is sufficient to overcome the trapping potential. Recently, it has been shown that these localized excitons are responsible for the observation of a pronounced single-photon emission from TMD monolayers [28-32] (Fig. 4b). The emitters often appear at the edges of the TMD samples and the spectral width of the emission is below 120 μeV in a free-standing WSe$_2$ monolayer [28].

**Interlayer excitons.** The fascinating exciton physics in atomically thin nanomaterials becomes even richer when considering that they can be vertically stacked to form VdW heterostructures [21]. Here, the strong Coulomb interaction gives rise to the formation of interlayer excitons, where the involved electrons and holes are located in different TMD layers (Fig. 2b) [22, 23, 71, 86-90]. After optical excitation of regular intralayer excitons, holes or electrons can tunnel to the other layer forming interlayer excitons. They show binding energies in the range of 100 meV [91, 92] making them stable at room temperature and robust against dissociation in applied electric fields. Depending on the spin and momentum of the involved states, these spatially separated electron-hole pairs can be either bright or dark. Note that the mismatch in the lattice constants of different layers results in a Moiré pattern in real space. The pattern can be controlled via the relative orientation angle between the layers and can have a significant impact on the optical response of the heterostructures [93, 94]. VdW heterostructures present an emerging field of research, where recently an increasing number of studies has been published, in particular demonstrating the appearance of interlayer excitons in PL spectra [22, 23, 86, 87]. Here, a pronounced additional resonance is observed below the energy of intralayer excitons. Its intensity is pronounced under weak excitation and clearly exceeds the one of the intralayer excitons (Fig. 4c). This is due to very efficient interlayer charge tunneling processes, which spatially separate electron-hole pairs generated by optical absorption in the individual monolayers. In time-resolved PL measurements (Fig. 4c), a very long-lived, spectrally narrow resonance was observed [95].

**Nonlinear optical response.** The light-matter interaction includes also higher-order nonlinear terms that play a particularly important role under intense optical excitation and are strongly enhanced by excitonic effects [96]. One important type of second-order optical nonlinearity is second-harmonic generation, or frequency doubling, which requires non-centrosymmetric crystal symmetry, and hence is absent in bulk or even-layered TMD crystals but is strong in monolayer or few odd-layered samples [97-99]. Its strength can be electrically controlled through exciton charging effects [100]. Polarization resolved second-harmonic measurements permit the detection of the crystallographic orientation [97-99], atomic edges and grain boundaries in polycrystalline films [101] as well as strains fields [102, 103]. Furthermore, several types of third-order nonlinear optical phenomena have been studied, including third-harmonic generation [104], four-wave mixing [105], saturable absorption [106], and others [107].

**OPTOELECTRONIC DEVICES**

**Electrically driven excitonic light emission.** Electroluminescence (EL) emission in a 2D semiconductor can occur as a result of both unipolar and bipolar currents. In the former case, emission is induced by only one type of carrier and has been ascribed to thermal population of exciton states as a result of Joule heating [108] (Fig. 5a) or exciton formation by impact excitation in a high electric field [109] (Fig. 5b). In the second case, electrons and holes are injected simultaneously into the 2D material, either in a lateral device configuration or in a VdW heterostructure, thus enabling exciton formation and radiative recombination. This was first accomplished in electrostatically defined p-n junctions using split-gate electrodes [36-38] (Fig. 5c, Fig. 6a). The same device structure allowed for the observation of interlayer exciton EL in a $MoSe_2/WSe_2$ heterobilayer [110]. Alternatively, electron-hole pairs can be injected in an ambipolar field-effect transistor using an ionic-liquid gate [111] or in a transient operation-mode by applying an AC voltage between the gate and the semiconductor [112]. Vertical VdW heterostructure light emitters, consisting of graphene electrodes, hBN tunnel barriers and a TMD monolayer semiconductor (Fig. 6b), have also been demonstrated [113]. Here, the hBN barriers prevent direct carrier tunneling between the graphene layers. In a related device structure [114, 115], minority carriers (in this case, holes) are injected from a graphene electrode through a single hBN tunnel barrier into an electrostatically (n-type) doped TMD sheet (Fig. 6c). This device structure also allowed for electrically driven quantum-light emission from localized exciton states, as confirmed by photon correlation measurements [114]. Another device concept is based on a mixed-dimensional silicon/TMD VdW heterojunction [116, 117], in which p-doped silicon serves as a hole injection layer into n-type $MoS_2$.

The EL quantum yield at room temperature is typically ~1% for both lateral and vertical device geometries [37, 112, 113, 115], which is of the same order of magnitude as the PL quantum yield at the equivalent excitation rate. The quantum yield currently is limited by non-radiative recombination and is expected to increase as higher quality samples become available. Light emission at current levels as low as ~100 pA has been demonstrated [38, 115]. It has further been shown that the emission is tunable between neutral excitons and trions [38, 113]. Although monolayers are usually preferred for electroluminescent devices because of their direct band gap, EL has also been observed in TMD multilayers by carrier redistribution from the indirect to the direct valleys in a high electric field [118], by indirect valley filling [119], or by the injection of hot carriers from a metal/TMD junction [120].

Spontaneous emission from a TMD monolayer can further be enhanced by evanescent coupling to an optical cavity [121, 122]. It has been reported that an optically pumped TMD semiconductor can provide sufficient gain to compensate for the losses in a high-Q cavity to achieve lasing [123-125]. Typical laser characteristics, such as a change in the slope of the output light intensity and emission linewidth narrowing, have indeed been observed, but it has also been argued [126] that reaching the lasing threshold is prevented by the broad TMD luminescence linewidth. Photon correlation measurement need yet to be carried out to validate the lasing action in TMDs.

**Photovoltaic solar cells.** Owing to their strong optical absorption in the solar spectral region [59], favorable band gaps for both single-junction and tandem cells [127] and high internal radiative efficiencies [42], TMDs provide an opportunity for the realization of ultrathin and ultralight photovoltaic devices. Theoretical power conversion efficiencies exceeding 25% have been predicted [127], indicating that these materials may become competitive or even outperform conventional photovoltaic technologies. Already early studies on bulk $MoS_2$ samples demonstrated the suitability of layered materials for photovoltaics [3]. More recently, lateral p-n junctions, realized by split-gate electrodes [36, 37, 128], and lateral metal/TMD Schottky junctions [129] have been investigated. Vertical junctions allow for easier scalability of the photoactive area and can be obtained by manual stacking or direct growth of different 2D semiconductors in a layered configuration. It has been shown that heterostructures composed of $MoS_2$ and $WSe_2$ form a type-II heterojunction [39-41], with the lowest-energy conduction band states in the $MoS_2$ layer and the highest-energy valence band states in $WSe_2$. Relaxation of photogenerated carriers, driven by the band offsets, then results in a charge transfer across the junction and a sizeable photovoltaic effect. Similar results have been obtained using other TMD semiconductor stacks with type-II band alignment [130-133]. Carrier extraction may either occur laterally (Fig. 7a), or vertically using transparent graphene electrodes (Fig. 7b). In an optimized $MoS_2/WSe_2$ device structure, an external quantum efficiency >50% and a (single-wavelength) power conversion efficiency of 3.4% have been achieved [134]. In addition, architectures have been explored in which chemical doping was applied to form a 2D homojunction in the same material.

Examples include plasma-induced p-doping of the top layers in an n-type $MoS_2$ multi-layer crystal [135] and mechanical stacking of n-type $MoS_2$:Fe and p-type $MoS_2$:Nb [136].

Despite their large optical absorption coefficient, TMD monolayers absorb only a fraction of the incident light. The absorption can be increased by using multilayers but is then limited to ~40% by surface reflection. A careful photonic design allows to overcome this limitation. One promising approach is to employ light trapping techniques, e.g. by placement of a TMD semiconductor with appropriate thickness on top of a metallic back reflector [137]. This resulted in >90% broadband absorption in an only 15 nm thick layer. The large exciton binding energy in 2D semiconductors poses another challenge for efficient power conversion. Although it has been shown that photogenerated excitons dissociate efficiently into free carriers by tunnel ionization in lateral device structures [138] or driven by the band offsets that exceed the exciton binding energy in type-II vertical junctions [139], the large binding energy nonetheless results in an open-circuit voltage penalty and hence reduced efficiency. This penalty is lower in multilayer structures than in monolayers. In addition, poor carrier transport properties [140] and inefficient carrier extraction at the contacts can lead to pile-up of photocarriers in the device and recombination losses. For all these reasons, power conversion efficiencies in TMD-based photovoltaic devices have as yet remained below the theoretical limit for their respective band gaps.

**Photodetection.** TMD-based photodetectors can roughly be classified by their detection mechanism into photovoltaic and photoconductive devices, as well as phototransistors. In addition, photo-excitation can heat the electron gas and produce a photovoltage due to the photo-thermoelectric (Seebeck) effect [141, 142]. The photovoltaic effect is based on the separation of photo-generated carriers, or dissociation of excitons, by built-in electric fields in lateral or vertical junctions, as described in the previous section. In photoconductive devices the photoexcited carriers increase the material's conductivity and an external bias is required to sense the conductance change [6, 34]. Fig. 7c shows a VdW heterostructure detector that can be operated in both regimes [143, 144]. No photovoltage is produced for a +10 V back-gate voltage, but the device conductance strongly increases under illumination (photoconductive mode). Other back-gate voltages break the device symmetry, due to doping of the bottom graphene sheet, and a photovoltage as well as zero-bias photocurrent are induced (photovoltaic mode). Typical responsivity and quantum efficiency values in these types of devices are in the order of ~$10^{-1}$ A/W and ~50%.

Although photovoltaic detectors give a response without bias, and thus without dark current, photoconductors are able to provide a photogain which can result in extraordinary high responsivities of more than $10^3$ A/W [35, 145]. The term photogain refers to the fact that the number of circuit electrons generated per absorbed photon is larger than one. In TMDs this typically results from capturing of one type of carriers (electrons or holes) into trap states that are present either in the underlying substrate or in the TMD itself. The former is exacerbated by the high surface-to-volume ratio in 2D materials and oxide surface treatments have been used to modify this effect. Also, a considerable contribution from extrinsic effects, such as charge transfer into surface bound water molecules or other adsorbates, has been observed [146] which leads to a conductivity change upon light exposure that persists for a long time after the light is switched off [147]. This phenomenon, called persistent photoconductivity, limits the practical usability of such devices because of the extremely slow response time and the sensitivity to the environment. It can be avoided by TMD encapsulation [148]. In a field-effect transistor, the trapped charge not only increases the channel conductance but also shifts the transistor's threshold voltage [146]. This mechanism is referred to as photogating. Low dark current can then be achieved by field-effect modulation to depletion. In general, higher responsivity values come at the cost of lower temporal bandwidth. This tradeoff can be controlled by the gate voltage, for example in Ref. [148] in the range $10$–$10^4$ A/W and $10$–$10^4$ ms. Because these devices may exhibit high dark currents, a proper assessment of their performance is not provided by the responsivity, but rather by the specific detectivity. Detectivity values up to $7.7 \times 10^{11}$ Jones have been reported [148]. An interesting approach to extend the operation range of TMD photodetectors towards longer wavelengths is the integration with sensitizing centers such as quantum dots [149, 150] (Fig. 7d). Incident light is absorbed in the dots, resulting in a charge transfer into the underlying TMD transistor which modifies the channel conductance. Responsivities of more than $10^5$ A/W and low dark currents have been achieved, which resulted in detectivity values as high as $5 \times 10^{12}$ Jones.

**Valley-optoelectronic devices.** The possibility of valley-selective excitation of excitons in TMD semiconductors provides the opportunity of using the valley degree-of-freedom as a new paradigm in data processing and transmission. To date, most work has focused on PL experiments, due to the relative simplicity of this technique, but it has been demonstrated that the valley polarization can also be detected electrically via the valley Hall effect [151] (Fig. 8a). Charge carriers in the K and K' valleys experience effective magnetic fields with opposite signs and are therefore deflected in opposite directions perpendicular to the current flow. The sign of the Hall voltage could be controlled by the helicity of the light, whereas no voltage was detected for linearly polarized light or in inversion-symmetric TMD bilayers. Other works demonstrated valley dependent photodetection in CVD grown $MoS_2$ [152], as well as the generation of spin-valley-coupled circular photogalvanic currents in $WSe_2$ [153]. In contrast to electrical fields that dissociate excitons and drive electrons and holes to opposite directions, giving rise to the valley Hall effect, statistical forces originating from temperature/chemical potential gradients drive electrons and holes in the same direction. This permitted the observation of the exciton Hall effect [154] by locally exciting a high density of valley-polarized excitons in a TMD monolayer and observing the thermally driven exciton diffusion using polarization-resolved PL imaging. Along with the longitudinal diffusion, a valley-selective transverse transport of excitons was observed (Fig. 8b).

Several studies have demonstrated that valley polarization can be electrically induced in a 2D semiconductor. This was first accomplished in a lateral $WSe_2$ p-n junction, defined by ionic-liquid gating [119]. The EL from this device was found to be circularly polarized, with the helicity depending on the current flow direction. The observation was attributed to different electron-hole overlaps in the K and K' valleys, that can be controlled by an in-plane electric field. Another work [155] employed spin-polarized charge carrier injection from a ferromagnetic electrode into a VdW heterostructure, resulting in valley polarization due to spin-valley locking in TMD semiconductors. The resulting valley polarization led to circularly polarized light emission that could be tuned by an external magnetic field (Fig. 8c). Valley polarization by spin injection has also been realized by using a magnetic semiconductor Ga(Mn)As [156].

**CHALLENGES**

In spite of the large number of studies on exciton physics in 2D semiconductors, the remarkably versatile excitonic landscape including bright, dark, localized, and interlayer excitonic states has partially remained literally in the dark. In particular, the relative spectral position of dark and bright states is still unclear and plays a crucial role for the optical response and the non-equilibrium dynamics in these materials. Furthermore, while exciton-phonon scattering has been well understood, the impact of exciton-exciton scattering warrants further investigation. Also, strategies to activate spin- and momentum-forbidden dark excitons are needed to exploit the full potential of possible exciton-based applications. On the application side, scaling of micrometer-sized proof-of-principle devices to macroscopic dimensions remains a challenge. Wafer-scale synthesis of uniform TMD layers has been achieved [157, 158], but the monolithic growth of VdW heterostructures is difficult. Another key issue is the lack of a controllable and reliable doping scheme, which detrimentally impacts the performance of devices in various ways. For example, it makes it difficult to create low-resistance ohmic contacts and carrier injection in TMDs is often obstructed by large Schottky barriers. In addition, non-radiative recombination, mediated by the high defect concentration in TMDs or by biexcitonic recombination, limits the performance of optoelectronic devices. While the former can be circumvented by employing higher quality material or chemical treatment, the latter restricts the range of useful carrier concentrations.

## ACKNOWLEDGEMENTS

The authors acknowledge support from the European Union (grant agreement No. 785219 Graphene Flagship), the Austrian Science Fund (START Y 539-N16), and the Swedish Research Council (VR).

## COMPETING INTERESTS

The authors declare no competing interests.

## AUTHOR CONTRIBUTIONS

Both authors contributed equally to the literature research and writing of the manuscript.

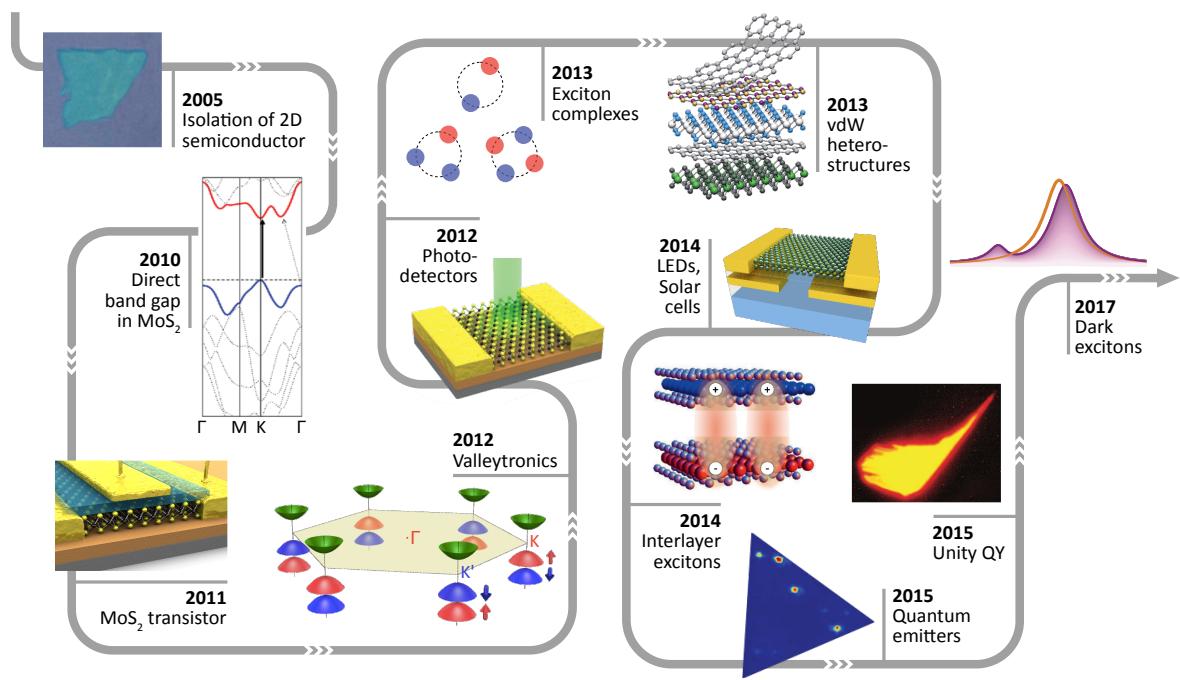

**Fig. 1** (Incomplete) history of TMD exciton optics and optoelectronics. Adapted with permission from ref. 5 (Copyright 2005 National Academy of Sciences), ref. 7 (Copyright 2010 American Chemical Society), ref. 16 (Copyright 2012 American Physical Society), ref. 21 (Copyright 2013 Springer Nature), ref. 24 (Copyright 2016 Springer Nature), ref. 30 (Copyright 2015 Springer Nature), ref. 35 (Copyright 2013 Springer Nature), ref. 36 (Copyright 2014 Springer Nature), and ref. 42 (Copyright 2015 American Association for the Advancement of Science). Image of $MoS_2$ transistor courtesy of A. Kis, EPFL.

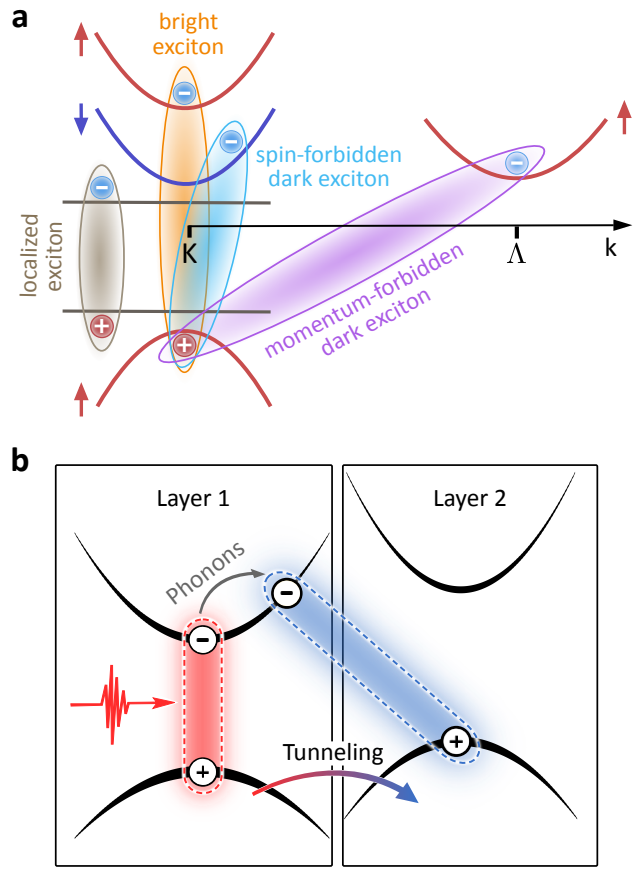

**Fig. 2** Different exciton types in atomically thin nanomaterials and related heterostructures. **a** Excitons are Coulomb-bound electron-hole pairs (ovals in the picture). Momentum-forbidden dark excitons consist of electrons and holes located at different valleys in the momentum space. Spin-forbidden dark excitons consist of electrons and holes with opposite spin. These states cannot be accessed by light due to the lack of required momentum transfer and spin-flip, respectively. Electrons and holes in localized excitons are trapped into an impurity-induced potential. **b** Interlayer excitons appear, where electrons and holes are located in different layers.

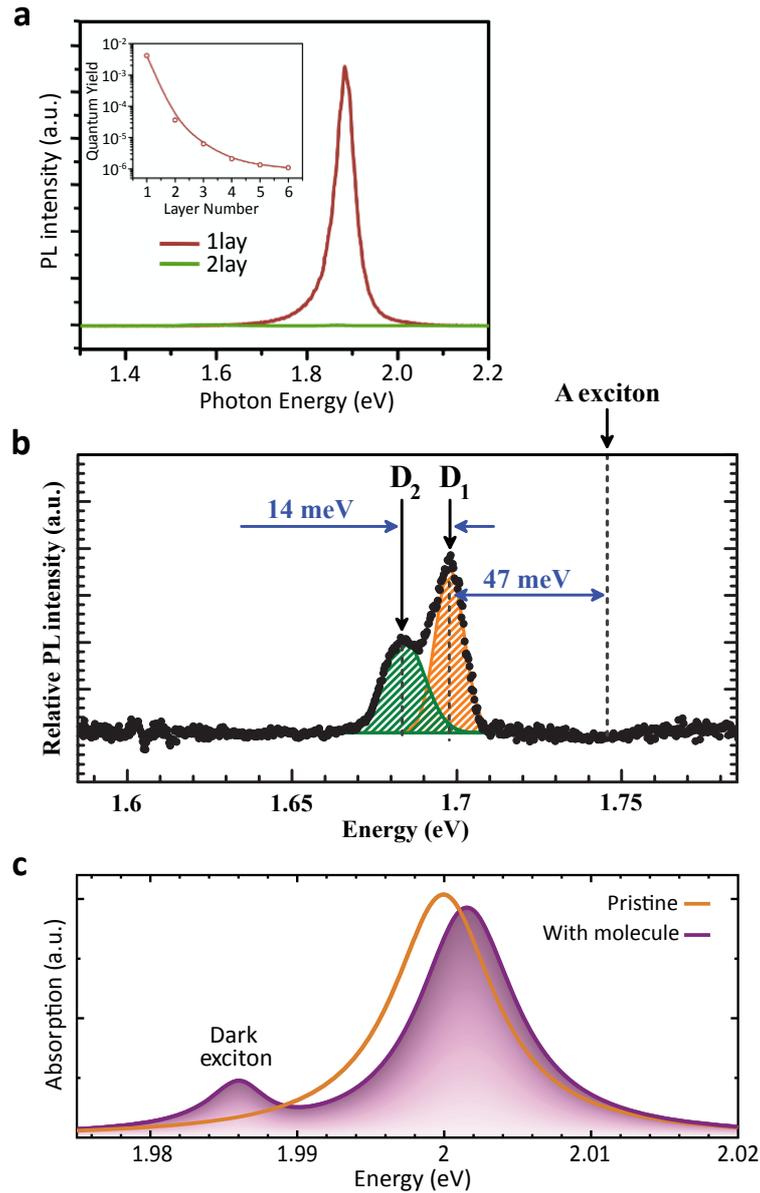

**Fig. 3** Photoluminescence spectra and dark exciton features. **a** Drastically enhanced layer-dependent PL intensity for MoS$_2$ monolayers confirming the transition to a direct-gap semiconductor. Adapted with permission from ref. 6 (Copyright 2010 American Physical Society). **b** Relative PL spectra of WS$_2$ monolayers, where the signal measured at the magnetic field $B = 14$ T was subtracted from the signal at zero field. The two additional resonances (D$_1$, D$_2$) below the A exciton can be ascribed to spin-forbidden dark excitons. Adapted from ref. 25. **c** Appearance of momentum-forbidden dark excitons in presence of molecules suggesting a new sensing method. Adapted from ref. 80.

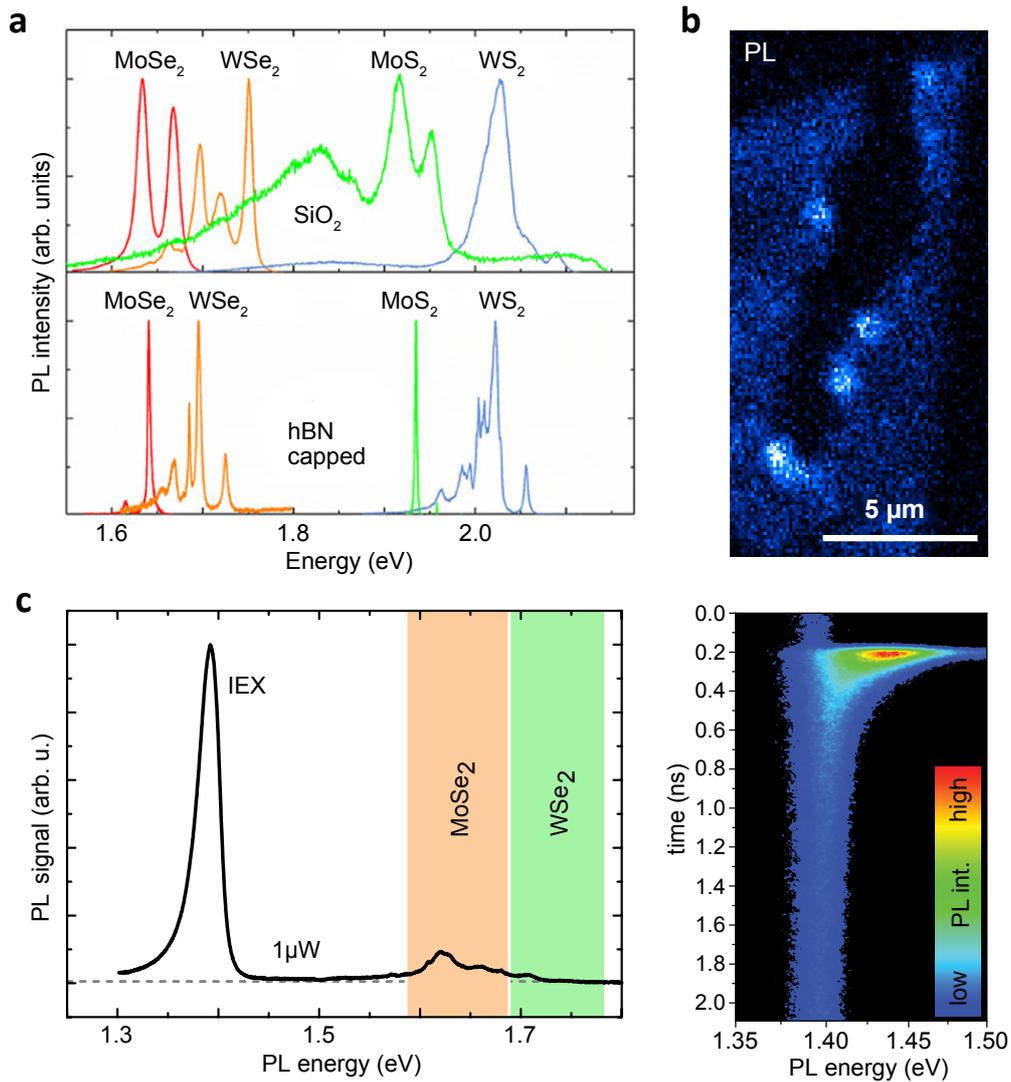

**Fig. 4** Localized and interlayer excitons. **a** PL spectra for different TMD monolayers at 4 K illustrating a series of resonances at low energies that can be ascribed to localized excitonic states (top). hBN-encapsulated TMDs show narrow linewidths (bottom). Adapted from ref. 85. **b** PL images of a WSe$_2$ monolayer scratched with a needle illustrating pronounced light emission from edges [28] (courtesy of R. Bratschitsch, Univ. Münster). **c** Energy- and time-resolved PL of a MoSe$_2$/WSe$_2$ heterostructure illustrating the appearance of strongly pronounced interlayer excitons. Adapted from ref. 86.

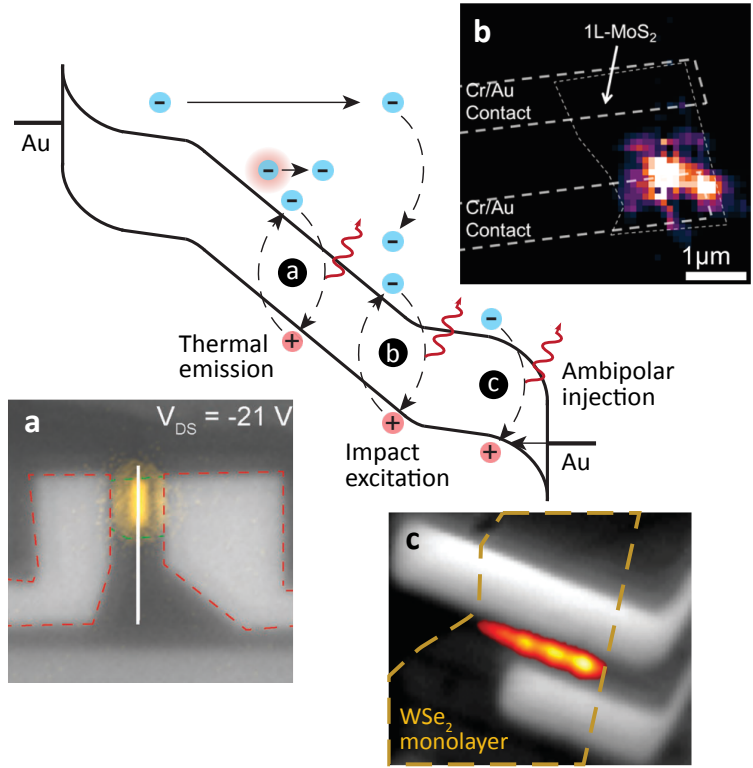

**Fig. 5** Electroluminescence mechanisms in TMD monolayers. **a** Thermal light emission, **b** impact excitation, and **c** electron-hole pair injection. Adapted with permission from ref. 108 (Copyright 2017 John Wiley and Sons), ref. 109 (Copyright 2010 American Chemical Society), and ref. 38 (Copyright 2014 Springer Nature), respectively.

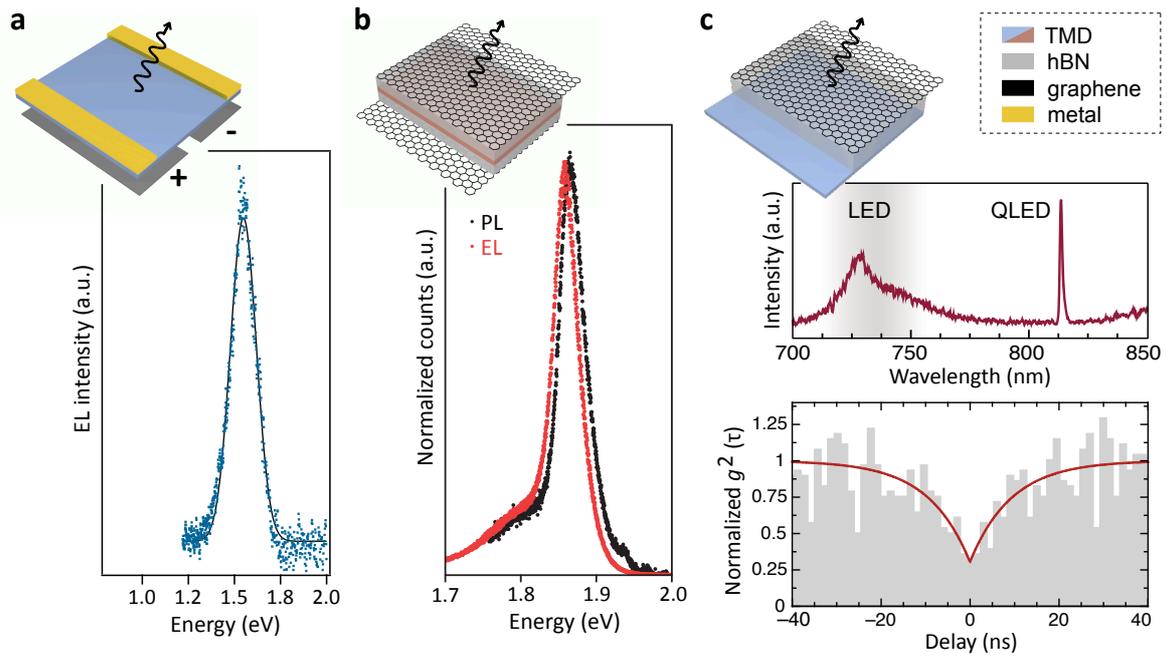

**Fig. 6** Electroluminescent devices. **a** EL from a lateral light-emitting diode with split-gate electrodes. **b** Graphene/hBN/TMD/hBN/graphene VdW heterostructure and comparison between EL and PL spectra. Adapted with permission from ref. 113 (Copyright 2015 Springer Nature). **c** Graphene/hBN/TMD device, showing a narrow emission line from a localized exciton state (top). Adapted from ref. 114. Photon correlation measurements of the EL show photon antibunching (bottom). The insets show schematic drawings of the corresponding device structures.

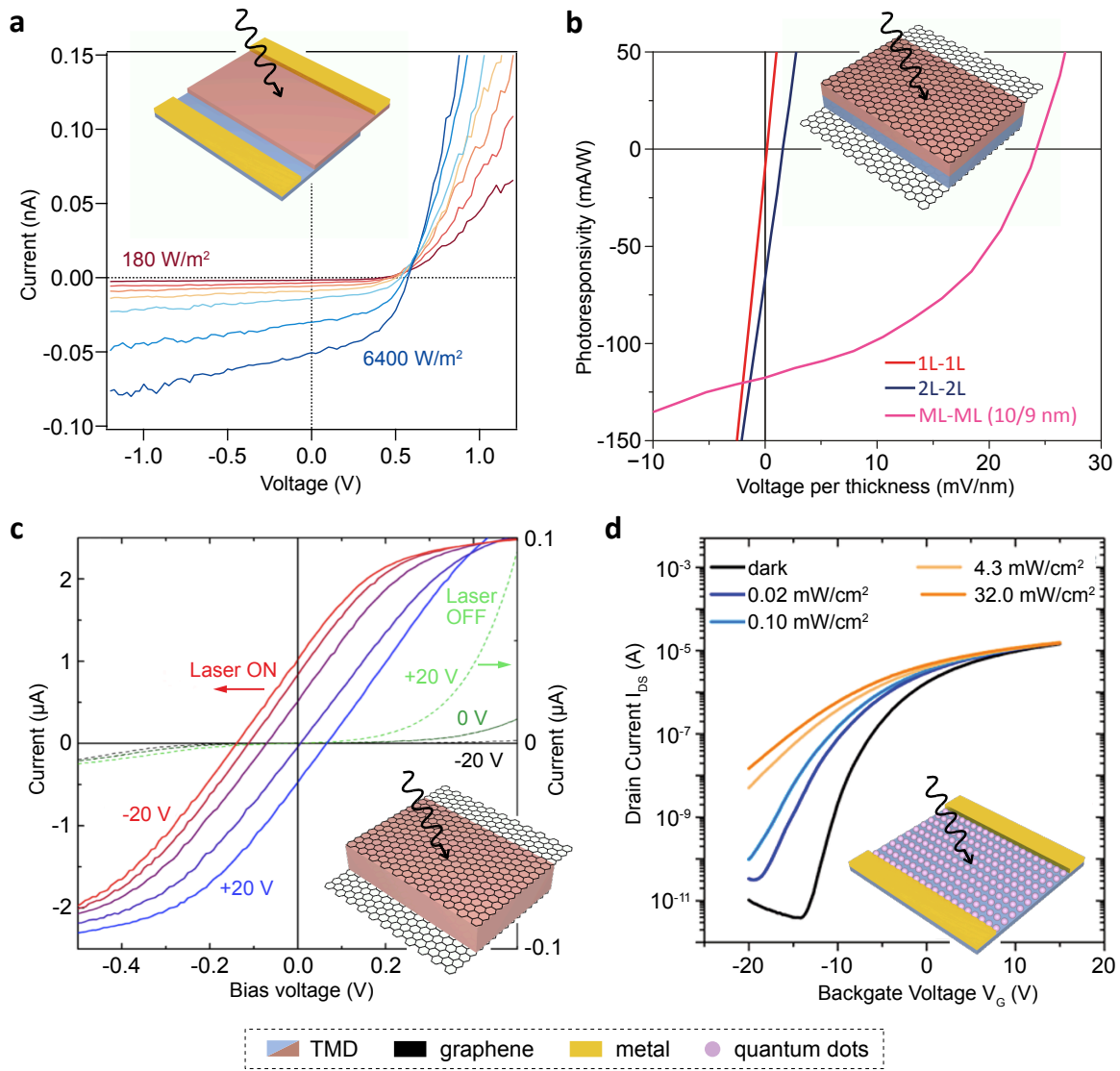

**Fig. 7** Photovoltaic solar cells and photodetectors. Photovoltaic effect in MoS$_2$/WSe$_2$ solar cells with **a** lateral and **b** vertical charge extraction. Adapted with permission from ref. 39 and ref. 40 (Copyright 2014 Springer Nature), respectively. **c** Graphene/TMD/graphene photodetector, with symmetry-breaking by a gate field. Adapted with permission from ref. 143 (Copyright 2013 American Association for the Advancement of Science). **d** Hybrid TMD/quantum dot photodetector with high photoconductive gain. Adapted with permission from ref. 150 (Copyright 2016 American Chemical Society). The insets show schematic drawings of the corresponding device structures.

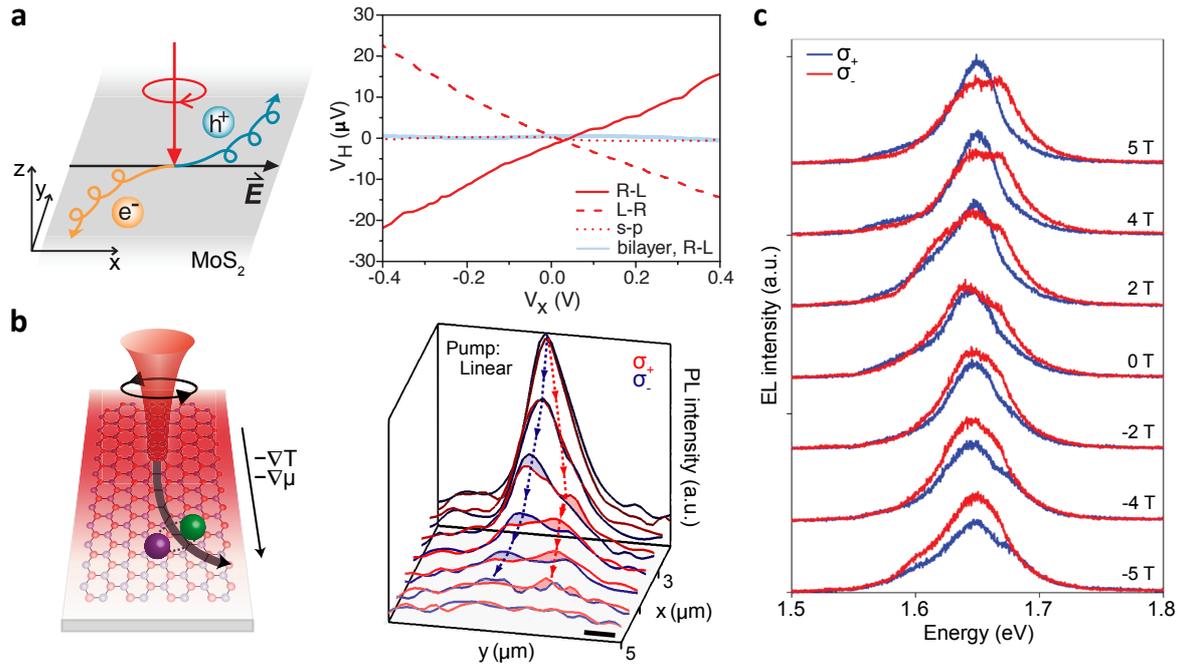

**Fig. 8** Valley-optoelectronics. **a** Experimental demonstration of the valley Hall effect. In a monolayer, the sign of the photovoltage produced in a Hall bar depends on the helicity of the light, whereas the effect is absent in a bilayer or for linear polarization. Adapted with permission from ref. 151 (Copyright 2014 American Association for the Advancement of Science). **b** Observation of the exciton Hall effect by thermally driven exciton diffusion and polarization-resolved PL imaging. Adapted with permission from ref. 154 (Copyright 2017 Springer Nature). **c** Spin-polarized charge carrier injection from a ferromagnetic electrode into a VdW heterostructure light emitter leads to circularly polarized light emission that can be tuned by an external magnetic field. Adapted from ref. 155.